\documentclass{ws-procs9x6}            
\begin{document}
\title{Wormholes Immersed in Rotating Matter}

\author{Christian Hoffmann$^1$,
Theodora Ioannidou$^2$,
Sarah Kahlen$^1$, 
Burkhard Kleihaus$^1$, 
Jutta Kunz$^1$}

\address{
$^1$
Institut f\"ur Physik, Universit\"at Oldenburg, Postfach 2503\\
D-26111 Oldenburg, Germany\\
E-mail: 
christian.hoffmann@uni-oldenburg.de;
sarah.kahlen@uni-oldenburg.de;
b.kleihaus@uni-oldenburg.de; 
jutta.kunz@uni-oldenburg.de\\
$^2$
Faculty of Civil Engineering,  School of Engineering\\
Aristotle University of Thessaloniki, 54249, Thessaloniki, Greece\\
E-mail:
ti3@auth.gr}

\begin{abstract}
We consider Ellis wormholes immersed in rotating matter in the form of an 
ordinary complex boson field. The resulting wormholes may possess full 
reflection symmetry with respect to the two asymptotically flat spacetime 
regions. However, there arise also wormhole solutions where the reflection 
symmetry is broken. The latter always appear in pairs. We analyse the 
properties of these rotating wormholes and show that their geometry may 
feature single throats or double throats. We also discuss the ergoregions 
and the lightring structure of these wormholes.
\end{abstract}


\bodymatter

\section{Introduction}\label{aba:sec1}

The non-trivial topology of wormholes requires the presence of exotic matter
in Einstein's General Relativity (see e.g.~the recent review
\cite{Lobo:2017oab} of the field).
Choosing a massless phantom (scalar) field for the exotic matter,
Ellis \cite{Ellis:1973yv,Ellis:1979bh} and Bronnikov \cite{Bronnikov:1973fh}
found static spherically symmetric wormhole solutions,
which connect two asymptotically flat regions of space-time.

Their rotating generalizations were first constructed perturbatively
for slow rotation
\cite{Kashargin:2007mm,Kashargin:2008pk}
and later numerically for rapid rotation
\cite{Kleihaus:2014dla}.
In these wormhole solutions the rotation of the throat
and thus the spacetime is 
achieved by an appropriate choice of the
boundary conditions. However, this results in the fact that
the two asymptotic regions are rotating with respect to
one another. Thus while both asymptotic regions are
asymptotically flat, the spacetime is not symmetric
with respect to reflection at the throat.

In order to obtain rotating wormholes that exhibit
a reflection symmetry at the throat, one can immerse
the throat inside rotating matter
\cite{Hoffmann:2017vkf,Hoffmann:2018oml}.
Then the rotation of the matter will drag the spacetime
and thus the throat of the wormhole.

\section{Wormholes Immersed in Rotating Matter}

A nice model to study such wormholes immersed in rotating matter
is obtained by adding an ordinary massive complex scalar field
to the real phantom scalar field into the action,
coupling both to gravity.
Without the phantom scalar field the model would yield
non-rotating and rotating boson stars. 
The combination of both scalar fields then allows for 
rotating wormholes immersed in scalar matter, 
that are reflection symmetric.
Moreover, a new type of non-symmetric wormholes emerges.
In the following we will briefly discuss the model
and present its rotating wormhole solutions and
analyze their properties
\cite{Hoffmann:2017vkf,Hoffmann:2018oml,Dzhunushaliev:2014bya,Hoffmann:2017jfs}.

\subsection{Theoretical Setting}

We consider the action $S$
\begin{equation}
S=\int \left[ \frac{1}{2 \kappa}{\cal R} + 
{\cal L}_{\rm bs} +{\cal  L}_{\rm ph} \right] \sqrt{-g}\  d^4x  
 \label{action}
\end{equation}
with the Einstein-Hilbert term,
the Lagrangian ${\cal L}_{\rm bs}$ of the complex scalar field $\Phi$
\begin{equation}
{\cal L}_{\rm bs} = 
-\frac{1}{2} g^{\mu\nu}\left( \partial_\mu\Phi^* \partial_\nu\Phi
                            + \partial_\nu\Phi^* \partial_\mu\Phi 
 \right) - m_{\rm bs}^2 |\Phi|^2  \ ,
\label{lphi}
\end{equation}
and the Lagrangian ${\cal L}_{\rm ph}$ of the phantom field $\Psi$,
\begin{equation}
 {\cal L}_{\rm ph} = \frac{1}{2}\partial_\mu \Psi\partial^\mu \Psi \ .
\label{lpsi}
\end{equation}

The field equations then consist of the Einstein equations
\begin{equation}
G_{\mu\nu}= {\cal R}_{\mu\nu}-\frac{1}{2}g_{\mu\nu}{\cal R} =  \kappa T_{\mu\nu}
\label{ee} 
\end{equation}
and the matter field equations
\begin{equation}
\nabla^\mu \nabla_\mu \Psi =0 \ 
\label{epsi} 
\end{equation}
and 
\begin{equation}
\nabla^\mu \nabla_\mu \Phi 
   = 
   m_{\rm bs}^2 \Phi \ .
\label{ephi} 
\end{equation}

An appropriate Ansatz for the metric is given by
\begin{equation}
ds^2 = -e^{f} dt^2 
    +e^{q-f}\left[e^b(d\eta^2 + h d\theta^2)+ h \sin^2\theta
    (d\varphi -\omega dt)^2\right] \ ,
\label{lineel}
\end{equation}
where  $f$, $q$, $b$ and $\omega$ are functions of
$\eta$ and $\theta$,
$h = \eta^2 +\eta_0^2$ with throat  parameter $\eta_0$,
and $\eta$ takes positive and negative
values, $-\infty< \eta < \infty$.
The ansatz for the rotating bosonic matter is taken as for boson stars 
\begin{equation}
\Phi(t,\eta,\theta, \varphi) 
  =  \phi (\eta,\theta) ~ e^{ i\omega_s t +  i n \varphi} \ ,   \label{ansatzp}
\end{equation}
where $\phi (\eta,\theta)$ is a real function,
$\omega_s$ is the boson frequency, $n$ is a rotational quantum number, 
and the ansatz for the phantom field $\Psi$ is simply
\begin{equation}
\Psi(t,\eta,\theta, \varphi) 
=  \psi (\eta,\theta) \ .
\label{ansatzph}
\end{equation}

The resulting set of six coupled non-linear partial differential equations
is then solved numerically subject to an appropriate set of 
boundary conditions in the two asymptotic
regions $\eta \to \pm \infty$, on the axis of rotation $\theta = 0$, and in the
equatorial plane $\theta = \pi/2$.

\subsection{Symmetric Wormholes}

\begin{figure}[t!]
\begin{center}
\mbox{\hspace{0.2cm}
{\hspace{-0.7cm}
\includegraphics[height=.238\textheight, angle =0]{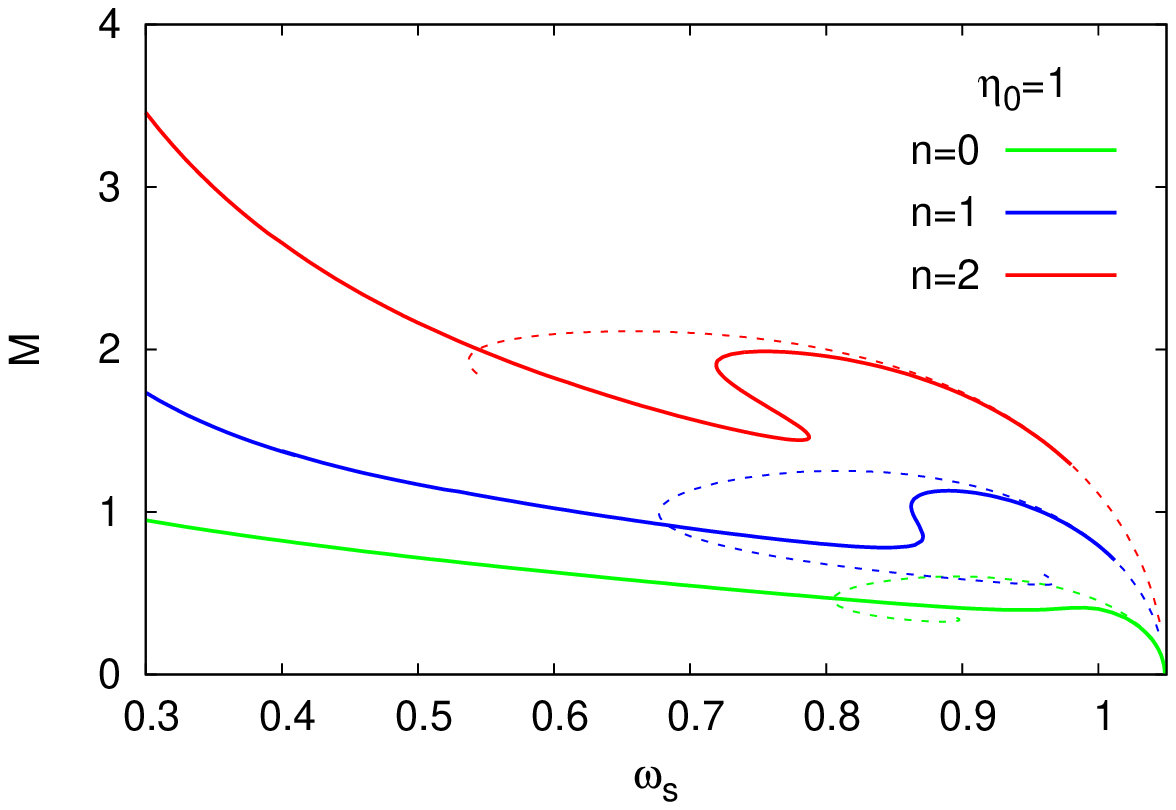}
}
{\hspace{-0.5cm}
\includegraphics[height=.238\textheight, angle =0]{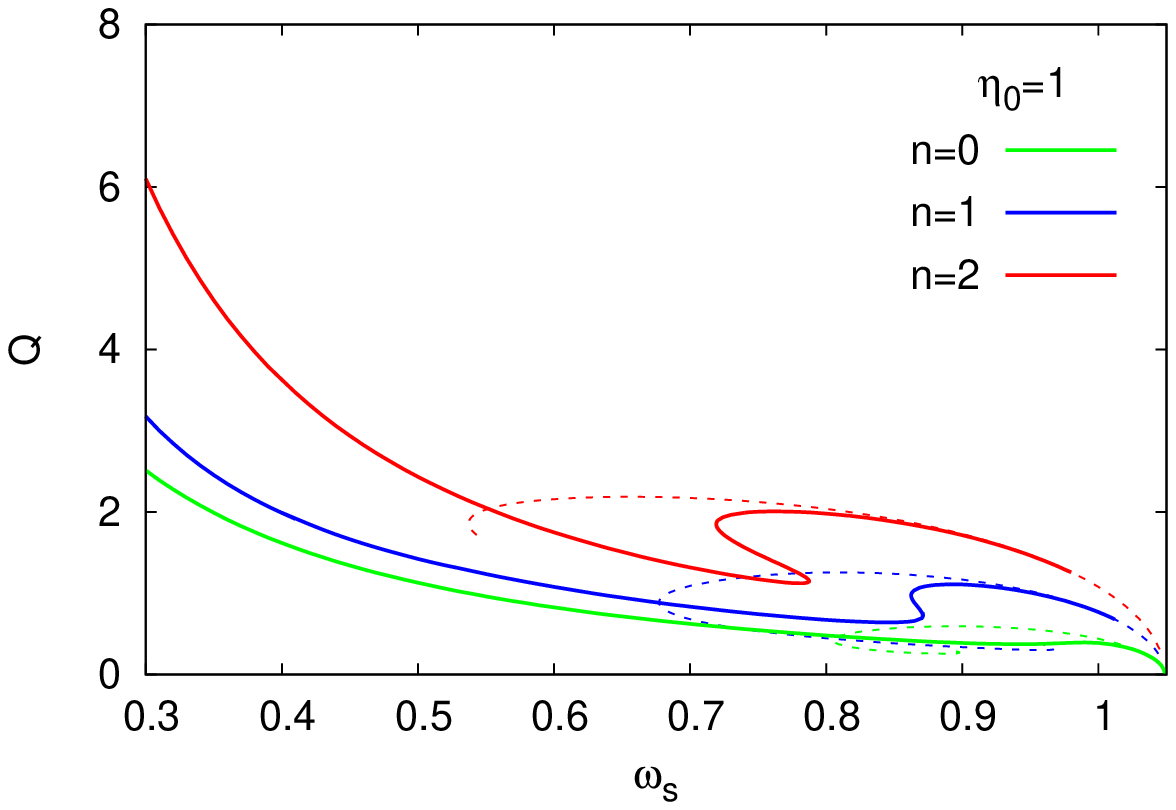}
}
}
\mbox{\hspace{0.2cm}
{\hspace{-0.7cm}
\includegraphics[height=.238\textheight, angle =0]{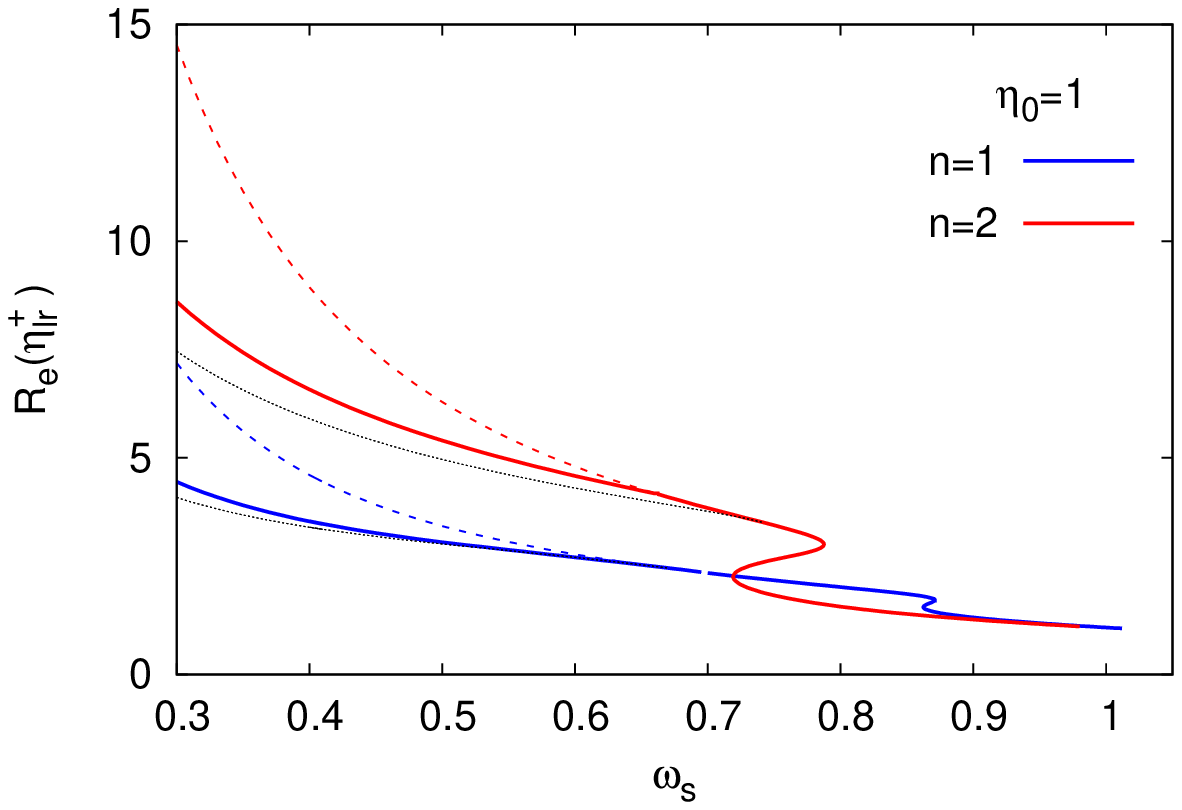}
}
{\hspace{-0.5cm}
\includegraphics[height=.238\textheight, angle =0]{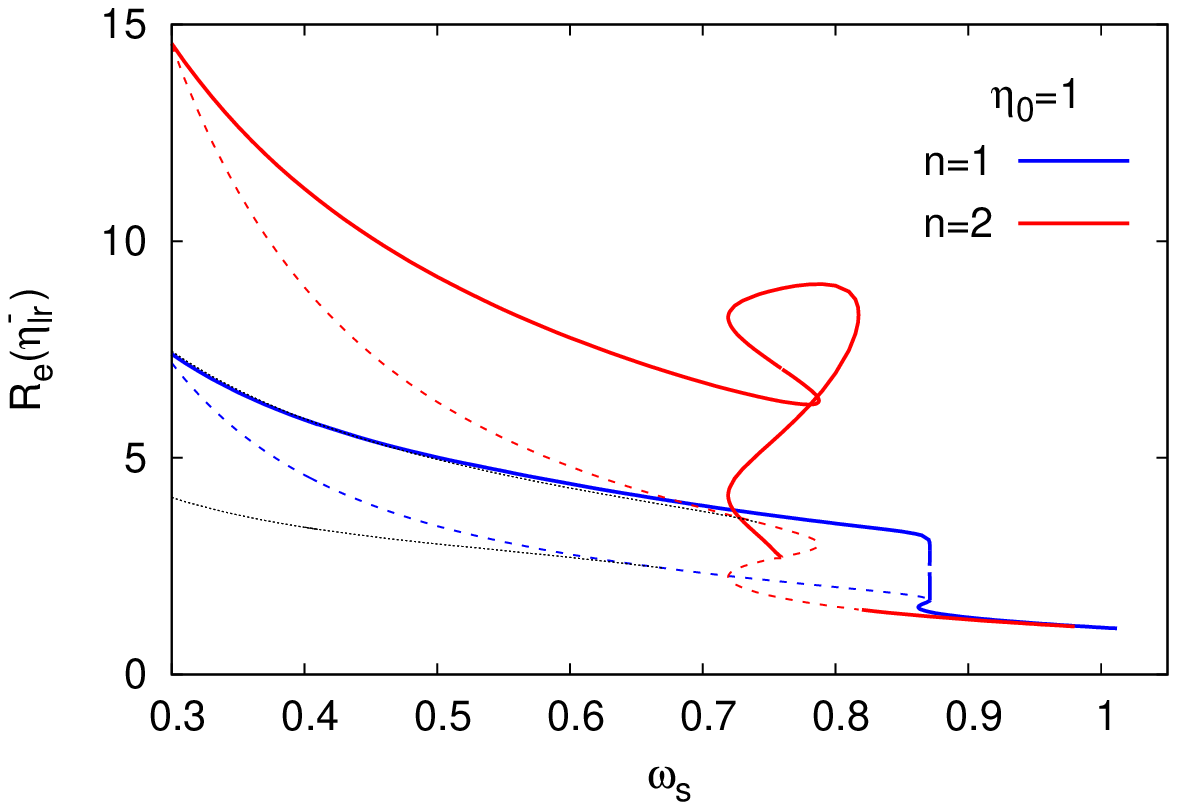}
}
}
\end{center}
\vspace{-0.5cm}
\caption{
Properties of symmetric wormhole solutions
(throat parameter $\eta_0=1$,
rotational quantum numbers $n=0$, 1, 2)
versus the boson frequency $\omega_s$:
(a) the mass $M$;
(b) the particle number $Q$;
the dashed lines indicate the respective boson star solutions;
(c) the corotation lightring 
circumferential radius $R_{\rm e}(\eta^+_{\rm lr})$;
(d) the counterrotation lightring 
circumferential radius $R_{\rm e}(\eta^-_{\rm lr})$;
also shown are the circumferential radii
of the ergosurfaces (black lines).
\label{Fig1}
}
\end{figure}

Let us now discuss the properties of the symmetric wormholes immersed
in bosonic matter.
In Fig.\ref{Fig1}a and b we show their mass $M$ and particle number $Q$
versus the boson frequency $\omega_s$ for a typical set
of such wormholes ($\eta_0=1$, $n=0$, 1, 2).
Their angular momentum $J$ is given by $J=nQ$.

Clearly, the domain of existence is limited by
a maximal value $\omega_{\rm max}=m_b$,
where a vacuum configuration with $M=0=Q$ is reached,
analogous to boson stars.
For large values of $\omega_s$
the global charges of the wormholes follow those of boson stars
(see the thin black lines in the figures).
However, for small $\omega_s$ the spiralling behaviour 
present in boson stars is basically lost.
In fact, the would-be spirals unwind
with respect to the frequency and continue to lower frequencies
(possibly reaching a singular configuration 
\cite{Dzhunushaliev:2014bya}).

We emphasize that these solutions satisfy the same boundary conditions
in both asymptotic regions. Thus in the case of rotation,
it is the complex scalar field with its finite rotational
quantum number $n$ that imposes the rotation on the configuration.
Then the rotation of the scalar field is reflected in the rotation
of the spacetime, leading to a rotating throat and frame dragging.
Not too surprisingly therefore a sufficiently fast rotation
will lead to ergoregions in the wormhole spacetimes.
The circumferential radii of the ergoregions
are exhibited by the black lines in Fig.~\ref{Fig1}c and d.

When we consider the geometry of the wormhole solutions
we realize that there arises a transition from ordinary single throat
wormholes to double throat wormholes with an equator in between,
as the frequency $\omega_s$ is decreased.
At the transition value the throat degenerates to an
inflection point, i.e., the circumferential radius at the center $\eta=0$
has vanishing first and second derivative.

Of interest are also the lightrings of these spacetimes,
as exhibited in Fig.~\ref{Fig1}c and d
for corotating and counterrotating photon orbits, respectively.
In particular, we show
their circumferential radii $R_{\rm e}(\eta^+_{\rm lr})$
and $R_{\rm e}(\eta^-_{\rm lr})$. 
We note that  a single lightring exists for large $\omega_s$.
For smaller $\omega_s$ two more lightrings emerge.
One lightring is always located at $\eta=0$
(i.e., at the throat or equator),
and the additional ones are located symmetrically w.r.t.~$\eta=0$.
In the $n=2$ case up to five lightrings of counterrotating massless
particles exist.

\subsection{Asymmetric Wormholes}

\begin{figure}[t!]
\begin{center}
\mbox{\hspace{0.2cm}
{\hspace{-0.7cm}
\includegraphics[height=.238\textheight, angle =0]{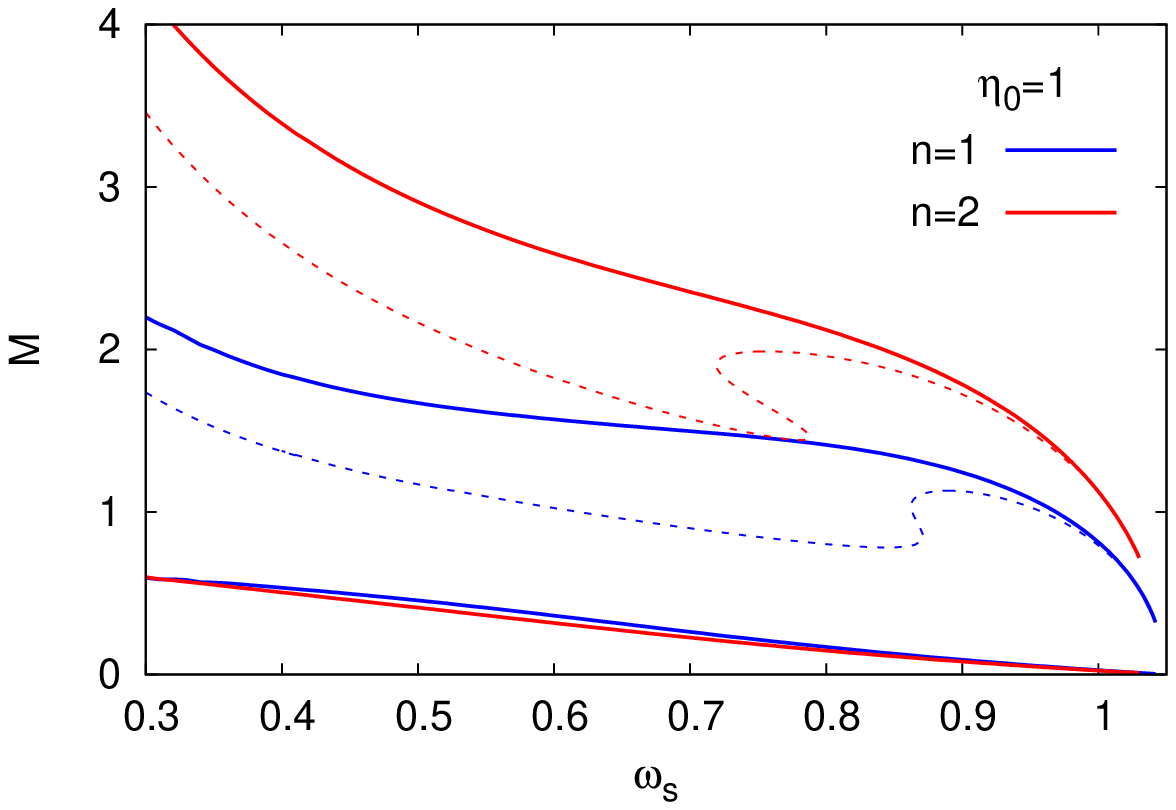}
}
{\hspace{-0.5cm}
\includegraphics[height=.238\textheight, angle =0]{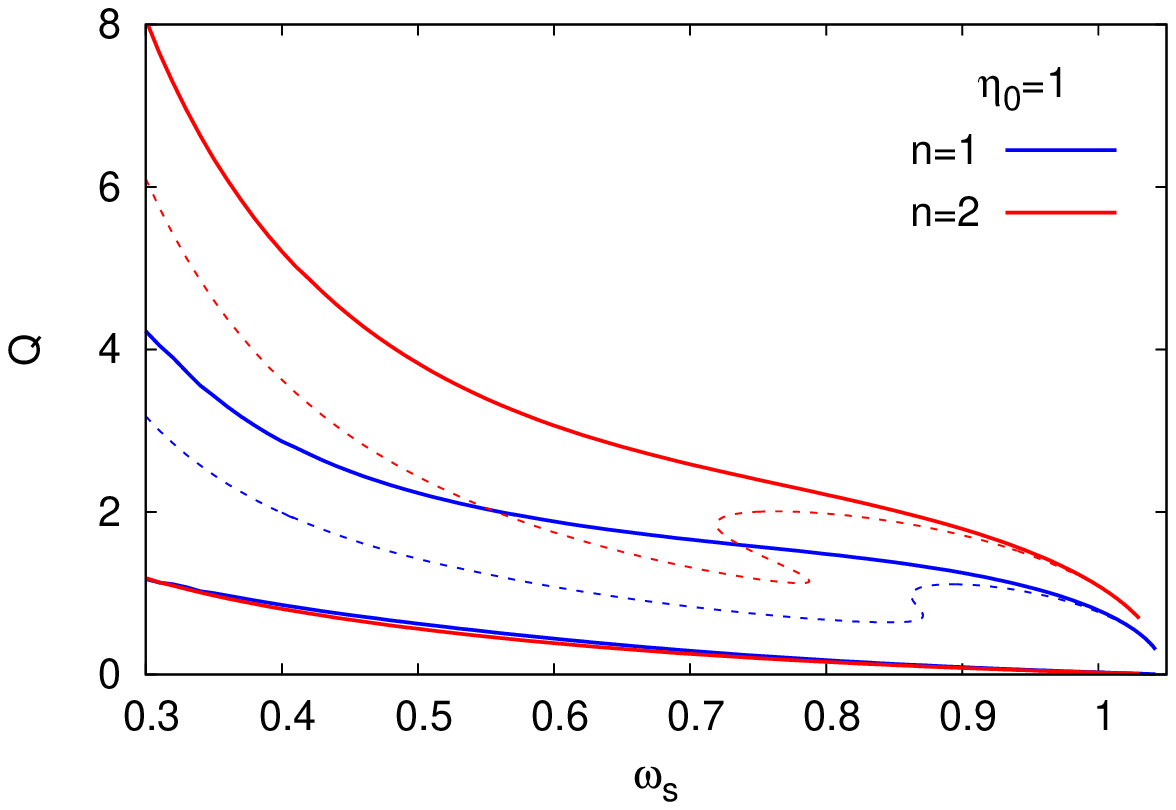}
}
}
\end{center}
\vspace{-0.5cm}
\caption{
Properties of asymmetric wormhole solutions
(throat parameter $\eta_0=1$,
rotational quantum numbers $n=1$, 2)
versus the boson frequency $\omega_s$:
(a) the mass $M$;
(b) the particle number $Q$;
the dashed lines indicate the respective symmetric solutions.
\label{Fig2}
}
\end{figure}

Let us now focus on another new aspect,
namely the presence of symmetric and asymmetric wormhole solutions
immersed in bosonic matter.
Whereas the field equations are symmetric with respect
to reflection of the radial coordinate at the center,
and the same boundary conditions are employed
in both asymptotically flat regions,
the solutions may, however, still be either symmetric or asymmetric
with respect to such a reflection. 
It is the non-trivial topology, which allows for asymmetric
solutions, as well.

Starting again the discussion with the global charges,
we exhibit the mass and the particle number of these asymmetric solutions
in Fig.~\ref{Fig2}, and compare to the corresponding symmetric solutions.
Because of the asymmetry the boson field is different in both
regions of the spacetime, resulting in different global
charges, read off asymptotically. Thus for a given $n$ there is
one curve for the symmetric wormholes, but there are two curves
for the asymmetric wormholes.
Note, that for these asymmetric solutions 
the angular momentum no longer satisfies $J=nQ$.

As in the case of the symmetric wormholes,
there arise double throat wormholes, but now the equator
will not reside at $\eta=0$, but instead arise somewhere
in one of the regions. Of course, for each asymmetric solution
there exists a second solution, obtained for $\eta \to -\eta$.
The asymmetric wormholes may also possess ergoregions
and multiple lightrings.

\section{Conclusions and Outlook}

\begin{figure}[t!]
\begin{center}
\mbox{\hspace{0.2cm}
{\hspace{-0.7cm}
\includegraphics[height=.238\textheight, angle =0]{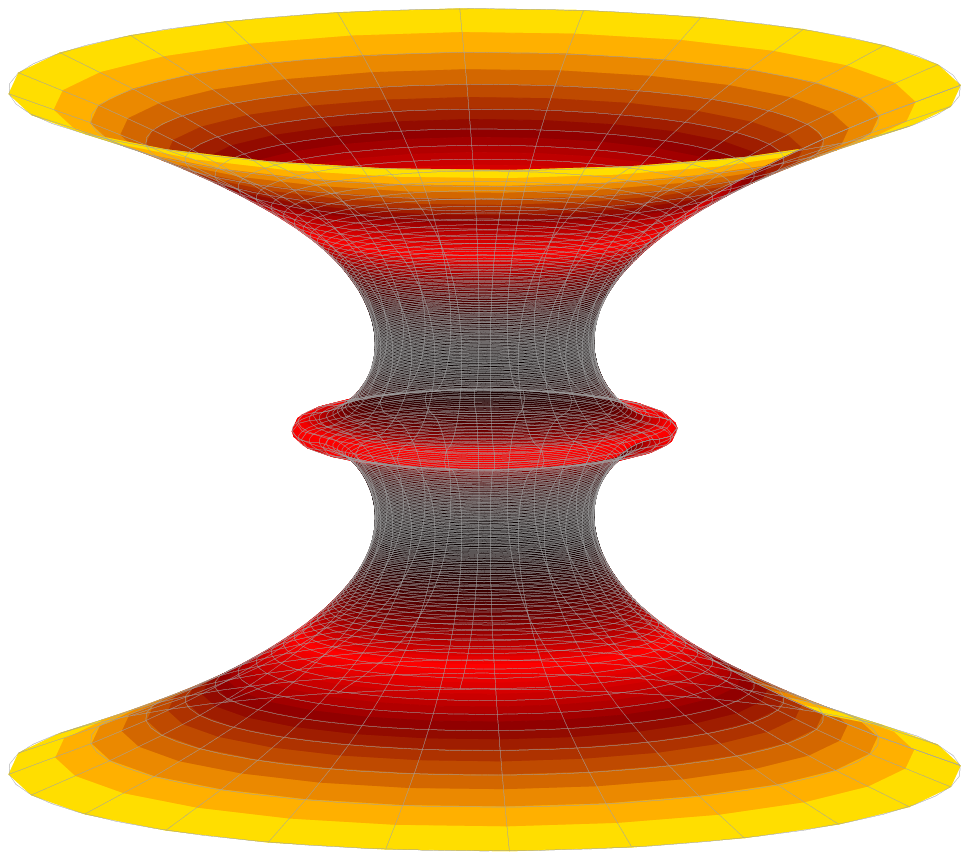}
}
{\hspace{-0.5cm}
\includegraphics[height=.238\textheight, angle =0]{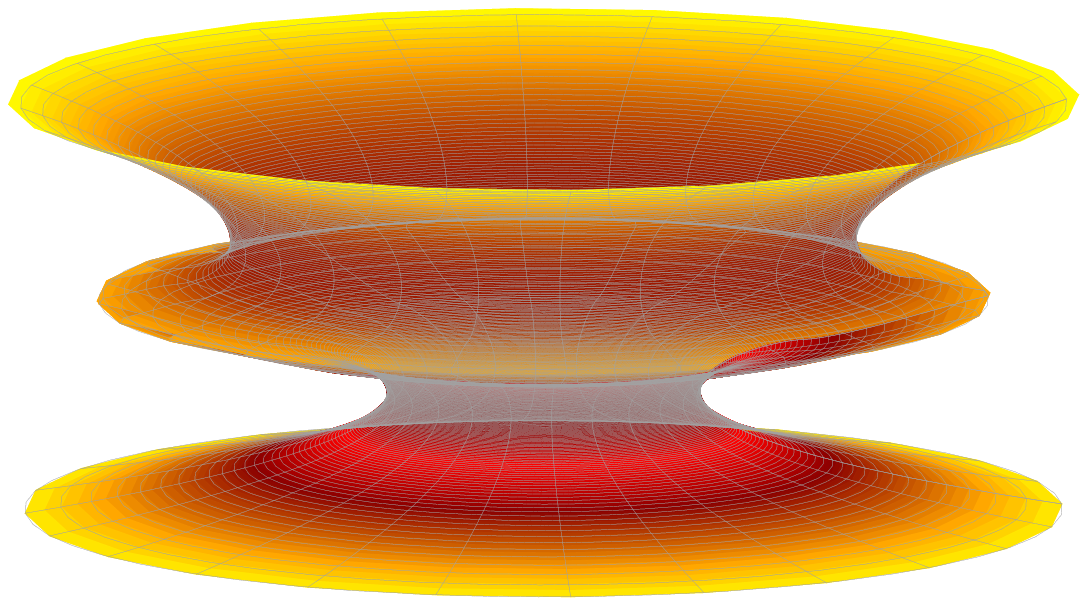}
}
}
\end{center}
\vspace{-0.5cm}
\caption{
Embeddings of symmetric (a) and asymmetric (b)
wormhole solutions.
\label{Fig3}
}
\end{figure}

We have obtained a new type of rotating wormhole by immersing
the throat in rotating bosonic matter,
adopting some features from boson stars.
We have studied various physical properties of these solutions,
like their global charges, their ergoregions, and their lightrings.
To conclude,
let us illustrate these new wormhole solutions
immersed in rotating matter,
via embeddings of a symmetric
and an asymmetric double throat wormhole in Fig.~\ref{Fig3}.

All these wormholes are based on General Relativity and therefore
need exotic matter for their existence. It will be interesting
to consider this new type of wormhole solutions in
generalized theories of gravity, which allow for
wormholes without the need for exotic matter.
Another point of interest will be the study of the stability
of these solutions.

\end{document}